\documentclass[11pt]{article}
\usepackage{fancybox,amsfonts,amsmath}
\usepackage[dvipdfm]{graphicx}
\usepackage{subfigure}
\usepackage{amsmath,amssymb,fancyhdr}

\def\firstpage{\hfill RUP-11-4}
\makeatletter
\def\ps@titlepage{%
	\@oddhead{\hfil\firstpage\hfil}%
}
\makeatother

\makeatletter
\renewcommand{\theequation}{\thesection.\arabic{equation}}
\@addtoreset{equation}{section}
\makeatother

\title{{\bfseries Minimal doubling fermion and hermiticity}}
\author{Syo Kamata and Hidekazu Tanaka\\
{\it Department of Physics, Rikkyo University, Tokyo 171-8501, Japan}\\
{\it E-mail:} {\ttfamily skamata@rikkyo.ac.jp, tanakah@rikkyo.ac.jp}}
\date{}

\begin{document}
\maketitle
\thispagestyle{titlepage}

\begin{center}
\Large{Abstract}
\end{center}

We analyze the lattice fermion kinetic term using PT symmetry, R-hermiticity, and $\gamma_{5}$-hermiticity. 
R-hermiticity is a condition for Hermite action and it is related to $\gamma_{5}$-hermiticity and PT symmetry. 
Assuming that a translation-invariant kinetic term with continuum and periodic function does not have PT symmetry, it can have R-hermiticity or $\gamma_{5}$-hermiticity. 
We prove that a kinetic term with continuum and periodic function that is PT symmetric does not reduce doublers. 
As a simple example, we analyze the two-dimensional two-flavor Gross–Neveu model with minimal doubling fermions. 
The minimal doubling fermions break PT symmetry and R-hermiticity, hence complex or non-Hermite coupling constants are caused by quantum correction.
\newpage

\section{Introduction} \mbox{}\indent
Lattice gauge theory is a powerful tool for revealing nonperturbative quark dynamics \cite{Wi1}. 
In this formulation, the space-time coordinate is discretized and physical variables are defined at sites and links. We can calculate observables in the strong coupling region using various techniques, e.g., high temperature expansion. 
Monte Carlo simulations are particularly effective methods to investigate nonperturbative physics and are currently being carried out.
As is well known, a naive lattice fermion has redundant physical degrees of freedom, “doublers”; this is called the “doubling problem”. We cannot remove doublers without sacrificing some symmetries or properties, because of the no-go theorem of Nilsen and Ninomiya \cite{NN1}.
To overcome this problem, many fermion formulations have been constructed, e.g., the Wilson fermion \cite{Wi1} and the
KS fermion \cite{KS1}.
 The application of lattice formulation to theories involving matter fields, e.g., quantum chromodynamics (QCD), is a major problem. 
In particular, exact chiral symmetry is important in analyzing nonperturbative QCD; however, this symmetry is incompatible with the removal of doublers.
In recent years, Creutz constructed an exact chiral symmetric lattice fermion \cite{C1}, and Borici fitted it to an orthogonal lattice \cite{B1}.
 A few decades ago, Karsten constructed a fermion formulation with the
same structure, but with a different action to the Creutz one \cite{K1}.
These fermions are called “minimal doubling fermions” \cite{C1}-\cite{BBTW2}. 
The minimal doubling fermions break (hyper-)cubic symmetry and some discrete symmetries, such as charge conjugation (C), parity transformation (P), time reflection (T), and so on. 
Many properties of the fermions have been analyzed in an orthogonal lattice \cite{K1}-\cite{CKM1} and hyperdiamond lattice \cite{BBTW2}.
In quantum theory, we must fine-tune some parameters to preserve these symmetries; however, it is difficult to adjust them generally.
In this paper, we analyze the translation-invariant, continuum, and periodic function lattice fermion kinetic term using
$\gamma_{5}$-hermiticity, R-hermiticity, and PT symmetry. 
These symmetries and hermiticities are related to each other. 
For example, assuming that a translation-invariant kinetic term with continuum and periodic function does not have PT symmetry, it can have R-hermiticity or
$\gamma_{5}$-hermiticity. 
R-hermiticity is a reality or Hermite condition for renormalized coupling constants perturbatively. We show that a PT-symmetric kinetic term cannot reduce doublers. 
As a simple example, we apply minimal doubling fermions that do not have PT symmetry or R-hermiticity to the two-dimensional N-flavor Gross–Neveu model and calculate renormalization group flows. 
In this flow, complex or non-Hermite coupling constants are caused by quantum correction.
This paper is organized as follows. 
In Sect \ref{sec.3cond}, we discuss the relationship between PT symmetry,
R-hermiticity, and $\gamma_{5}$-hermiticity. 
In Sect. \ref{sec.con}, we conclude and summarize the paper.

\section{$\gamma_{5}$-hermiticity, R-hermiticity and PT symmetry} \label{sec.3cond} \mbox{}\indent
In this section, we define $\gamma_{5}$-hermiticity, R-hermiticity and PT symmetry in lattice fermion formulation and show how they restrict a kinetic term.
$\gamma_{5}$-hermiticity is closely related to sign problem, i.e., the fermion determinant is not a positive value, and it is sometimes used as an Hermite condition.
R-hermiticity is a classical Hermite condition, which is used in e.g., Ref. \cite{P1}.
In quantum theory, we will show that this condition restricts effective coupling constants to real values in perturbation.
PT symmetry is important for a fermion kinetic term and doublers.
We will discuss these issues in detail below.

For a concrete discussion, we will focus on only kinetic terms in four dimensions.
We can easily extend the following discussion to even dimensions.
We define a translation-invariant kinetic term in momentum space as follows (the lattice space $a=1$):
\begin{eqnarray}
S &=& \int \frac{d^{4} k}{(2 \pi)^{4}} \bar{\psi}(-k) D(k) \psi(k),
\end{eqnarray}
with
\begin{eqnarray}
D(k) &=& \sum_{\mu=1}^{4} f_{\mu}(k) \gamma_{\mu}, \label{eq.f_mu}
\end{eqnarray}
where $f_{\mu}(k)$ are complex numbers in general and $f(k)_{\mu} \rightarrow i k_{\mu}$ in the classical continuum limit.

We define $\gamma_{5}$-hermiticity, R-hermiticity and PT symmetry as follows
\footnote{In following discussion, we can use C symmetry instead of PT symmetry because of CPT theorem.}:
\begin{eqnarray}
\mbox{$\gamma_{5}$-hermiticity} &:& D(k) = \gamma_{5} D ^{\dagger} (k) \gamma_{5},\\
\mbox{R-Hermiticiy} &:& D(k) = D ^{\dagger} (-k), \\
\mbox{PT symmetry} &:& D(k) = \gamma_{5} D (-k) \gamma_{5}.
\end{eqnarray}
These conditions are not independent each other.
We can easily deduce that, if a kinetic term satisfies two of the three conditions, the other condition is automatically satisfied, and this fact is a sufficient condition.
If $f_{\mu}(k)$ is pure imaginary, $\gamma_{5}$-hermiticity assures an anti-Hermite condition for the kinetic term and a real positive fermion determinant.
R-hermiticity is also used as an Hermite condition, e.g., in Ref.\cite{P1}; however, it is not well-defined because the forward-derivative kinetic term, $D_{\mathrm{fd}}(k)=\sum_{\mu}\left( e^{ik_{\mu}}-1 \right) \gamma_{\mu}$, satisfies this condition.

Hence, we will show that R-hermiticity is a condition for real effective coupling constants in perturbation.
We assume that a fermion kinetic term has R-hermiticity and that the effective coupling constants have the following form:
\begin{eqnarray}
g_{eff} &=& g_{0} + \sum_{n=1}^{\infty} I^{(n)},
\end{eqnarray}
with
\begin{eqnarray}
I^{(n)} &=& \int_{-\pi}^{\pi} \prod_{i=1}^{r}  \frac{d^{4} k_{i} }{(2 \pi)^{4}} \cdot
I^{(n)}_{\alpha_{1} \beta_{1} \cdots \alpha_{r} \beta_{r}}(-k_{1}, \cdots, -k_{r}) \cdot \prod_{j=1}^{r} S_{\alpha_{j} \beta_{j}}(k_{j}), \label{eq.I}
\end{eqnarray}
where $g_{0}$ is a real bare coupling constant whereas $g_{\mathrm{eff}}$ is an effective coupling constant.
$S_{\alpha \beta}(k)$ is a fermion propagator and $I^{(n)}$ is the $n$-loop quantum effect, which is constructed from $r$-fermion propagators.
If Hermite conjugate acts on the second term on the r.h.s. of Eq.(\ref{eq.I}), the effective parameter is real if the following condition is satisfied;
\begin{eqnarray}
I^{(n)}_{\alpha_{1} \beta_{1} \cdots \alpha_{r} \beta_{r}}(-k_{1}, \cdots, -k_{r}) = I^{(n)\dagger}_{\beta_{1} \alpha_{1} \cdots \beta_{r} \alpha_{r}}(k_{1}, \cdots, k_{r}). \label{eq.I_he}
\end{eqnarray}
This equation is the Hermite condition for $I^{(n)}$.
If the action is constructed from Hermite terms except a fermion kinetic term, this equation is satisfied.
Therefore  R-Hermiticiy is a reality or Hermite condition for coupling constants as long as Eq.(\ref{eq.I_he}) is satisfied.

\begin{figure}[t]
\begin{center}
\begin{tabular}{c c}
\subfigure[$g^{2}=0$]{\includegraphics[width=80mm]{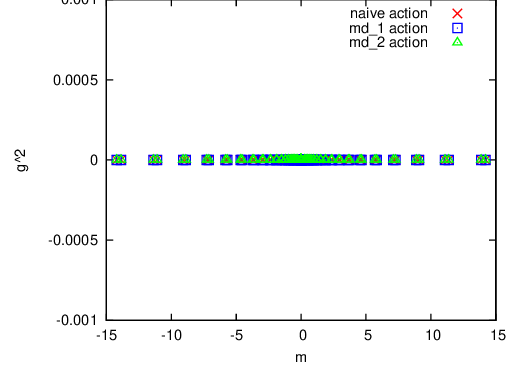}}
\subfigure[$g^{2}=0.2$]{\includegraphics[width=80mm]{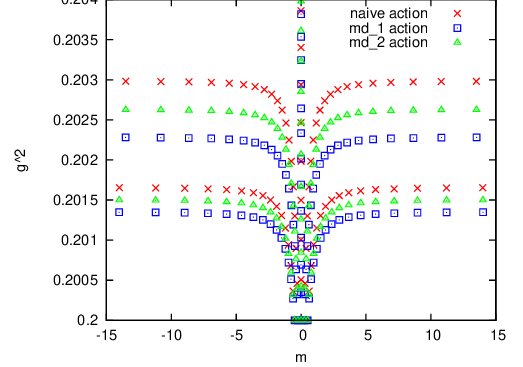}}\\
\subfigure[$g^{2}=0.4$]{\includegraphics[width=80mm]{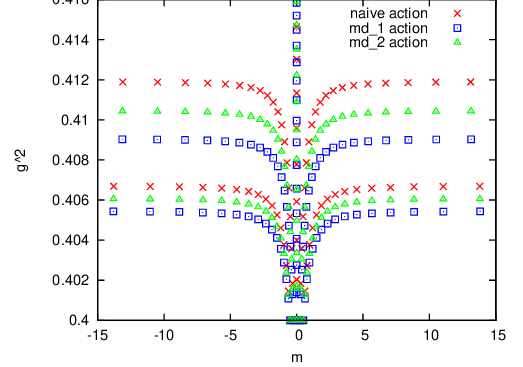}}
\end{tabular}
\end{center}
  \caption{The RGFs of the NA and MDAs. The initial parameters are $m_{11}=m_{22}=m_{33}=m_{44}=0,\pm 0.25,\pm 0.5,g^{2}=(a) 0, (b) 0.2, (c) 0.4$. The RGFs run from the initial conditions toward infrared, which the $g^{2}$ are increasing.} 
  \label{fig.g_n01234}
\end{figure}
\begin{figure}[t]
\begin{center}
\begin{tabular}{c c}
\subfigure[$D_{\mathrm{md}1}$]{\includegraphics[width=80mm]{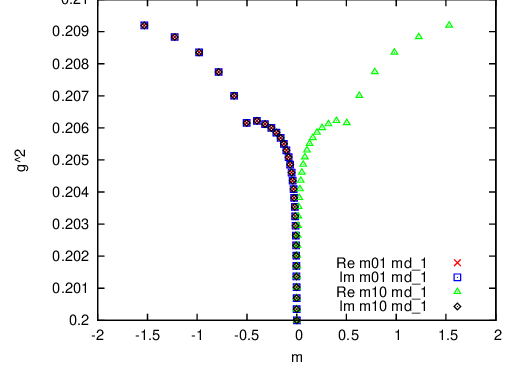}}
\subfigure[$D_{\mathrm{md}2}$]{\includegraphics[width=80mm]{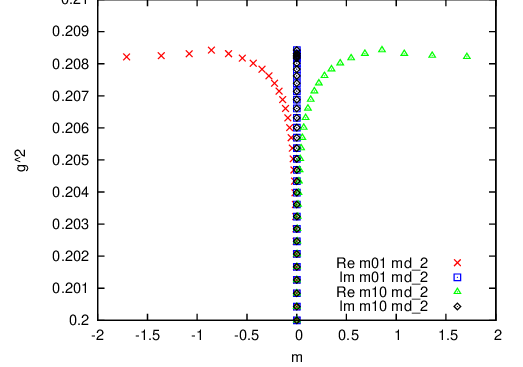}}
\end{tabular}
\end{center}
 \caption{The off-diagonal mass and coupling constant of the MDAs, (a)$D_{\mathrm{md}1}$, (b)$D_{\mathrm{md}2}$. The initial parameters are $m_{11}=m_{22}=m_{33}=m_{44}=0, g^{2}=0.2$. The RGFs run from the initial conditions toward infrared which the $g^{2}$ are increasing. The RGFs have irregular forms and the off-diagonal mass components show non-hermiticity.}
\label{fig.mass_md12}
\end{figure}
\begin{figure}[t]
\begin{center}
\begin{tabular}{c c}
\subfigure[$D_{\mathrm{md}1}$]{\includegraphics[width=80mm]{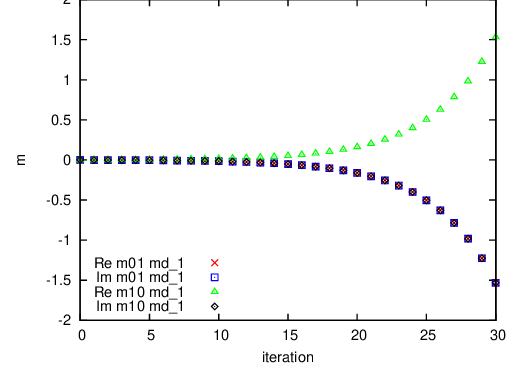}}
\subfigure[$D_{\mathrm{md}2}$]{\includegraphics[width=80mm]{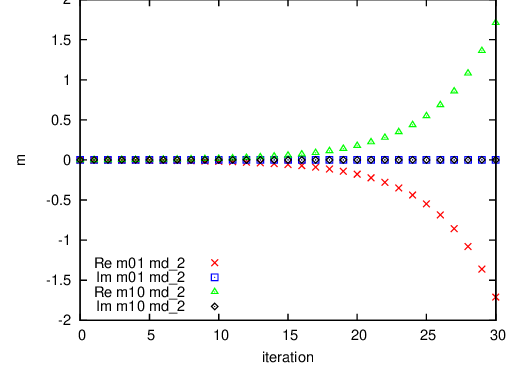}}
\end{tabular}
\end{center}
 \caption{The off-diagonal mass and iteration of the MDAs, (a)$D_{\mathrm{md1}}$, (b)$D_{\mathrm{md2}}$. The initial parameters are $m_{11}=m_{22}=m_{33}=m_{44}=0, g^{2}=0.2$. The off-diagonal mass components are generated and have non-hermiticity.}
 \label{fig.m_ite_md12}
\end{figure}

Figures.1-3 show renormalization grop flows(RGFs) of the Gross-Neveu model in two dimensions using naive action(NA) and minimal doubling actions(MDAs).
This is the simplest model for visualizing complex or non-Hermite coupling constants caused by quantum correction.
We define MDAs and Gross-Neveu model and explain how to calculate Wilsonian RGFs in Appendixes \ref{sec.MDA}, \ref{sec.NGN}, and \ref{sec.wil_RGF} respectively.
Because the MDAs have only $\gamma_{5}$-hermiticity, the mass term has an off-diagonal or complex quantum correction, which is proportional to a $\gamma$ matrix.
A more complicate example is given in Ref.\cite{CCWW1}.
Using the Wilsonian method \cite{WK1}, we calculate numerically the RGFs for the mass and coupling constant starting from the trivial fixed point, $m=g^{2}=0$. 
In the case of the MDAs, we use doublers as the different flavor fermions,
and, in the case of the NA, we use only two poles, $\tilde{p} = (0,0)$ and $(\pi,\pi)$.
We represent the spinor indices explicitly, and we distinguish $0,1$ from $2,3$ as different flavors.
We assume that high-frequency modes of fields $\psi(1<|k|)$, $\bar{\psi}(1<|k|)$, and $\sigma(1<|k|)$ are not effective, and we neglect their contributions.
We choose the initial conditions for the mass to be $m_{00}=m_{11}=m_{22}=m_{33}=0,\pm 0.25, \pm 0.5$, and for the coupling constant to be $g^{2}=0,0.2,0.4$. 
We set the off-diagonal mass components equal to zero in all cases.
We will calculate numerically the one-loop quantum effects and RGFs, which run from the initial conditions
\footnote{To estimate integrating part of the one-loop calculation,  we used the sectional measurement method.
We chose the division length to be $\Delta p_{\mu} = 0.01$. The error is $O(0.01^{2})$ by one iteration.}.
In our calculation we define $\gamma$ matrices as follows:
\begin{eqnarray}
\gamma_{1} = \left(
\begin{array}{ccc}
0 & -i \\ 
i & 0
\end{array}
\right),\,\,\,\,
\gamma_{4} = \left(
\begin{array}{ccc}
0 & 1 \\ 
1 & 0
\end{array}
\right).
\end{eqnarray}

The results are shown in Fig.\ref{fig.g_n01234}.
The RGFs of the NA and MDAs have similar forms and the difference between each value is $O(10^{-3})$.
In MDA cases, however, off-diagonal mass components are generated by the RGFs, except the initial value, which is a trivial fixed point, $m=0,g^{2}=0$.
We show this fact in Fig.\ref{fig.mass_md12} and \ref{fig.m_ite_md12} with an initial condition of $m=0,g^{2}=0.2$.
Figures.\ref{fig.mass_md12}(a) and (b) show the RGFs of $D_{\mathrm{md1}}(p)$ and $D_{\mathrm{md2}}(p)$, respectively.
Figures.\ref{fig.m_ite_md12}(a) and (b) show the relationship between the off-diagonal mass components and iterations using $D_{\mathrm{md1}}$ and $D_{\mathrm{md2}}$, respectively.
Though the off-diagonal mass components amplify as the flows approach IR because of the scaling effect, they do not break chiral $\mathbf{Z}_{4}$ symmetry
\footnote{These generated couplings are essentially different from mass.
They do not break chiral symmetry because $\Delta m \propto i \gamma_{\mu}$.
These terms couple  $\bar{\psi}_{L}$ to $\psi_{L}$ and $\bar{\psi}_{R}$ to $\psi_{R}$. }.
These $\Delta m$ are not always complex but off-diagonal.
We can choose a $\gamma$ matrix representation, which diagonalizes one matrix, e.g.,
$
\gamma_{1} = \left(
\begin{array}{ccc}
1 & 0 \\ 
0 & -1 
\end{array}
\right),\,\,\,\,
\gamma_{4} = \left(
\begin{array}{ccc}
0 & -i \\ 
i & 0
\end{array}
\right).
$
In this representation, shifted mass is complex, 
$
\Delta m \propto i \gamma_{1} =\left(
\begin{array}{ccc}
i & 0 \\ 
0 & -i
\end{array}
\right).
$

In principle, these terms can be canceled by counterterms; therefore, we can fine-tune the perturbation \cite{CCWW1}.
However, nonperturbative analysis is difficult and this problem must be solved in future work. 

Next, we will show that PT symmetry is always broken if we add extra kinetic terms to a NA to reduce it to doublers
\footnote{This argument has been discussed in Ref.\cite{BBTW1}, although not mathematically.}.
\\ \\
{\bf Statement.} \\ 
{\it In even dimensions, a PT-symmetric kinetic term with assumed periodicity and continuity function always has equal to or more than $2^d$ poles}.
\\ \\
{\it Proof.} \\
For simplicity, we also assume translation invariance
\footnote{We can similarly arrive at same statement without translation invariance.
In the case of non-translation invariance, a kinetic term has two-momenta dependence $D(k,p)$, and at least $4^{d}$ doublers appear.}.
A general $2 \pi$ periodic and continuum $D(k)$ has the following form:
\begin{eqnarray}
D(k) = \sum_{\mu,\nu=1}^{d} \sum_{n \in \mathbf{N}^{d} }^{\infty}  \left[\left( A_{\mu\nu}(n) +i B_{\mu\nu}(n) \right)  \cos (n_{\nu} k_{\nu})
 +  \left( C_{\mu\nu}(n) +i D_{\mu\nu}(n) \right) \sin(n_{\nu}  k_{\nu})  + E_{\mu\nu} \right] \gamma_{\mu},
\end{eqnarray}
where $A_{\mu\nu}(n),B_{\mu\nu}(n),C_{\mu\nu}(n),D_{\mu\nu}(n)$ are real constants and $E_{\mu\nu}$ are complex constants.
From PT symmetry,
\begin{eqnarray}
&& A_{\mu\nu}(n) = B_{\mu\nu}(n)= E_{\mu\nu} =0, \,\,\,\, \mbox{for all} \,\,\,\, \mu,\nu, n.
\end{eqnarray}

The $D(k)$ always has two poles at $k=0$ and $\pi$ for each dimension.
Therefore $D(k)$ has equal to or more than $2^{d}$ poles.
\\

This statement means that we cannot reduce the number of doublers using PT-symmetric kinetic terms
\footnote{We cannot apply this statement to the non-$\gamma_{\mu}$ linear case, e.g., the Wilson fermion.}.
In a numerical simulation context, $\gamma_{5}$-hermiticity is a very important condition to avoid the sign problem.
Assuming translation invariance, R-hermiticity is not satisfied if $D(k)$ satisfies $\gamma_{5}$-hermiticity but not PT symmetry.
Therefore, the effective parameters have explicit non-hermiticity.

In the process of rewriting from Minkowskian to Euclidean, Hermite fermion kinetic terms transmute to anti-Hermite ones.
In the Minkowski formulation, we forbid non-Hermite or complex couplings using the Hermite condition.
In contrast, the definition of ``Hermite" in Euclidean space is ambiguous.
Though some MDAs have reflection symmetry or reflection positivity, which are equal to the Hermite conjugate or unitarity in Minkowski space, 
these conditions do not properly have non-hermiticity.
Similarly, $\gamma_{5}$-hermiticity is commonly used as an Hermite condition, but we cannot forbid non-Hermite or complex couplings directly. 
A kinetic term that reduces the number of doublers allows the possibility of generating these anti-Hermite effective coupling constants.
R-hermiticity is a criterion to remove non-hermiticity.

\section{Conclusion} \label{sec.con} \mbox{}\indent
We have analyzed the translation-invariant, continuum and periodic function lattice fermion kinetic term using $\gamma_{5}$-hermiticity, R-hermiticity and PT symmetry.
These conditions are not independent, because satisfying two of the three conditions is a sufficient condition for the other condition.
However, it is not a necessary condition.
Additionally we have suggested that R-hermiticity is a condition for removing non-hermiticity or complex couplings.

We have proved that the PT-symmetric kinetic term does not reduce doublers.
Because minimal doubling fermions have only $\gamma_{5}$-hermiticity it generates a renormalized non-Hermite or complex mass by quantum correction.
As a simple example of non-R-hermiticity, we visualize the complex coupling constant using one-loop Wilsonian renormalization group flows of the two-flavor Gross-Neveu model in two dimensions.

\subsection*{Acknowledgement} \mbox{}\indent
S.K. thanks M. Hashi, T. Kimura, T. Misumi, T. Noumi, and H. Suzuki for helpful discussions and comments.
This work was partially supported by the Research Center for Measurement in Advanced Science at Rikkyo University.

\newpage

\section*{Appendix} \label{sec.app}
\renewcommand{\theequation}{\thesubsection.\arabic{equation}}

\appendix
\def\thesubsection{\Alph{subsection}}

\subsection{Minimal doubling fermion} \label{sec.MDA} \mbox{}\indent
In this section, we will briefly review minimal doubling fermion actions briefly \cite{C1}-\cite{BBTW2}.
To analyze the quantum correction using two-dimensional GN model, we discuss minimal doubling fermion properties in two-dimensions here.

We define the kinetic terms of naive action(NA) and two minimal doubling actions(MDAs) in two-dimensional momentum space as follows:
\begin{eqnarray}
S_{\mathrm{kin}} &=& \int \frac{d^{2}p }{(2 \pi)^{2}} \bar{\psi}(-p) D(p) \psi(p),
\end{eqnarray}
where the subscript ``$\mathrm{kin}$" means a kinetic term, and

\begin{eqnarray}
D(p) &=& \left\{
\begin{array}{llll}
 \sum_{\mu=1,4} i \sin p_{\mu} \gamma_{\mu} & \equiv D_{\mathrm{n}}(p)  \\
 i ( \sin p_{1} + \cos p_{4} -1 )  \gamma_{1} + i ( \sin p_{4} + \cos p_{1} -1 )  \gamma_{4} & \equiv D_{\mathrm{md1}}(p) \\
 i (\sin p_{1} + \cos p_{4} -1) \gamma_{1} + i \sin p_{4}  \gamma_{4} & \equiv D_{\mathrm{md2}}(p) 
\end{array}.
\right.\nonumber \\ \label{eq.D(p)}
\end{eqnarray}
We fix a value of lattice space, $a=1$, from now on.
The subscripts $1,4$ mean the space and time components respectively.
$D_{\mathrm{md1}}$ and $D_{\mathrm{md2}}$ are called the ``twisted ordering action" and the ``dropped twisted ordering action" respectively
\footnote{
We have another choice of $D_{\mathrm{md2}}$ action, $D_{\mathrm{md2}}(p) = i (\sin p_{4} + \cos p_{1} -1) \gamma_{4} + i \sin p_{1}  \gamma_{1}$. This action does not have CP and T symmetry but has CT and P symmetries. In addition, this action does not have reflection symmetry, or reflection positivity. 
We can apply the same argument in this paper to another $D_{\mathrm{md2}}$ \cite{CM1}
}.

In two dimensions, the NA has four zero-modes and the MDAs have two, which appear in the following momenta:
\begin{eqnarray}
D_{\mathrm{n}} &:& \tilde{p}=(0,0),\,(0,\pi),\,(\pi,0) \, \mathrm{and}\,(\pi,\pi),\nonumber \\
D_{\mathrm{md1}} &:& \tilde{p}=(0,0) \, \mathrm{and}\,(\pi/2,\pi/2),\\ \label{eq.pole}
D_{\mathrm{md2}} &:& \tilde{p}=(0,0)\, \mathrm{and}\,(0,\pi).\nonumber
\end{eqnarray}
The doublers appear around each zero-mode as $D(p) = D(\tilde{p}+q)$ with $D(\tilde{p})=0$, and they contribute observables.
In the case of the NA, the half doublers have the same chirality and the others have opposite.

In cases of the MDAs, they have opposite chirality to each other.
The NA and MDAs have $\gamma_{5}$-hermiticity:
\begin{eqnarray}
\gamma_{5}D(p)\gamma_{5}=D^{\dagger}(p).
\end{eqnarray}
For massless fermions, they also have chiral symmetry:
\begin{eqnarray}
\gamma_{5}D(p)+D(p)\gamma_{5}=0.
\end{eqnarray}
The MDAs violate (hyper-)cubic symmetry and some discrete symmetries.
We define charge conjugation(C), parity transformation(P), time reflection(T), and their combinational transformation laws acting on a fermion kinetic term
\footnote{We can apply the same laws to four-dimensional theory, using $p_{1} \rightarrow \mathbf{p}$ instead.}:
\begin{eqnarray}
\mathrm{C} &:& D(p) \,\, \rightarrow \,\, -CD^{\top}C^{-1}(-p)\nonumber \\
\mathrm{P} &:& D(p) \,\, \rightarrow \,\, \gamma_{4} D(-p_{1},p_{4}) \gamma_{4} \nonumber \\
\mathrm{T} &:& D(p) \,\, \rightarrow \,\,
  \gamma_{1} D(p_{1},-p_{4}) \gamma_{1} \nonumber \\
\mathrm{CP} &:& D(p) \,\, \rightarrow \,\, - \gamma_{4} CD^{\top}C^{-1}(p_{1},-p_{4}) \gamma_{4} \\ \label{eq.dis_trf} 
\mathrm{CT} &:& D(p) \,\, \rightarrow \,\,
 -\gamma_{1} C D^{\top} C^{-1} (-p_{1},p_{4}) \gamma_{1} \nonumber \\
\mathrm{PT} &:& D(p) \,\, \rightarrow \,\, 
 \gamma_{5} D(-p) \gamma_{5}\nonumber \\
\mathrm{CPT} &:&  D(p)
\,\, \rightarrow -\gamma_{5} C D^{\top} (-p) C^{-1} \gamma_{5} \,\, \nonumber
\end{eqnarray}
We present these symmetric properties of the NA and MDAs in Table 1
\footnote{``T" represents site and link reflection in lattice space.
In the case of $D_{\mathrm{md2}}$, it has link reflection positivity \cite{P2}.}.

\begin{table}[t]
\begin{center}
\caption{Discrete symmetry for the NA and the MDAs}
\begin{tabular}{|c||c|c|c|c|c|c|c|}
 \hline
 & C & P & T & CP & CT & PT & CPT \\ \hline \hline
 naive & $\bigcirc$ & $\bigcirc$ & $\bigcirc$ & $\bigcirc$ & $\bigcirc$ & $\bigcirc$ & $\bigcirc$ \\ \hline
 md1 & $\times$ & $\times$ & $\times$ & $\times$ & $\times$ & $\times$ & $\bigcirc$ \\ \hline
 md2 & $\times$ & $\times$ & $\bigcirc$ & $\bigcirc$ & $\times$ & $\times$ & $\bigcirc$ \\ 
 \hline
\end{tabular}
\end{center}
\label{tbl.dis_prop}
\end{table}

\subsection{The $N$-flavor Gross-Neveu model and renormalization group flow in two dimensions} \label{sec.NGN} \mbox{}\indent
In this section, we describe the $N$-flavor Gross-Neveu(GN) model \cite{GN1} and calculate Wilsonian renormalization group flows(RGFs) using the NA and MDAs numerically.
Firstly we will review the $N$-flavor GN model and then we will calculate the RGFs.

We define the continuum Euclidean Lagrangian of the $N$-flavor GN model as follows: 
\begin{eqnarray}
\mathcal{L}_{\mathrm{GN}} &=& \bar{\psi} \left( \partial \cdot \gamma +m \right) \psi - \frac{g^{2}}{2N} \left( \bar{\psi} \psi \right) ^{2},
\end{eqnarray}
where $m$ is a fermion mass and $g$ is a coupling constant of the four-Fermi interaction. 
We omit flavor indices if we do not have to write them explicitly: $\bar{\psi}\psi \equiv \sum_{i=1}^{N} \bar{\psi}_{i} \psi_{i}$, where ``$i$" means flavor degrees of freedom.

This Lagrangian has $U(1)$ symmetry:
\begin{eqnarray}
\psi &\rightarrow& e^{i \theta} \psi, \nonumber \\
\bar{\psi} &\rightarrow& \bar{\psi} e^{-i \theta}.
\end{eqnarray}
In the case of massless fermions, this Lagrangian has chiral $\mathbf{Z}_{4}$ symmetry:
\begin{eqnarray}
\psi &\rightarrow& \left( i \gamma_{5} \right)^{n}\psi,\nonumber \\
\bar{\psi} &\rightarrow& \bar{\psi} \left( i \gamma_{5} \right)^n. \,\,\,\,\,\,\,\,(n=0,1,2,3)
\end{eqnarray}
In the case of massive fermions, chiral $\mathbf{Z}_{4}$ symmetry reduces to chiral $\mathbf{Z}_{2}$ symmetry($n=0,2$).
In addition, if all flavors have the same masses, it has $SU(N)_{F}$ symmetry:
\begin{eqnarray}
\psi_{i} &\rightarrow& U_{ij}\psi_{j},\nonumber \\\bar{\psi}_{i} &\rightarrow& \bar{\psi}_{j} U^{\dagger}_{ji} , \,\,\,\,\,\,\,\,(U \in SU(N))
\end{eqnarray}

It is convenient to redefine the GN action using an auxiliary scalar field $\sigma$ instead of $(\bar{\psi}\psi)$:
\begin{eqnarray}
\mathcal{L}_{\mathrm{GN}} &=& \bar{\psi} \left( \partial \cdot \gamma +m \right) \psi + \frac{N}{2} \sigma^{2} + g \sigma \bar{\psi} \psi .\label{eq.GN}
\end{eqnarray}
According to this manipulation, we can obtain the action which involves Yukawa interaction instead of four-Fermi interaction.
According to a perturbative calculation, the GN model has asymptotic freedom \cite{W2,TV1}.

\subsection{The Wilsonian renormalization group} \label{sec.wil_RGF} \mbox{}\indent
In this appendix, we review a method to calculate the Wilsonian renormalization group flow in the case of the GN model in two dimensions \cite{WK1}.

We define the partition function of the GN model in momentum space
\footnote{We omit the subscript, which indicates flavor.}:
\begin{eqnarray}
Z &=& \int D\sigma D\psi D\bar{\psi} \exp(-S_{\mathrm{GN}}),
\end{eqnarray}
with
\begin{eqnarray}
S_{\mathrm{GN}} &=& \int_{0 < |p| <1} \frac{d^{2}p}{(2 \pi)^{2}} \mathcal{L}_{\mathrm{GN}},
\end{eqnarray}
where 
\begin{eqnarray}
D\sigma = \prod_{0< |k| <1} d \sigma(k), \,\,\,\, D\psi = \prod_{0< |k| <1} d \psi(k), \,\,\,\,D\bar{\psi} = \prod_{0< |k| <1} d \bar{\psi}(k),
\end{eqnarray}
and $\mathcal{L}_{\mathrm{GN}}$ is given in Eq.(\ref{eq.GN}).
We can treat $N$ as a mass parameter of the auxiliary field $\sigma$.
Here we assume that the high-frequency modes have already integrated and they effectively do not contribute.
Then we split the field configurations as follows:
\begin{eqnarray}
\sigma(p) = \sigma_{l}(p) + \sigma_{h}(p),
\end{eqnarray}
where
\begin{eqnarray}
\sigma_{l}(p) &=& \sigma(p) \,\,\,\, \hbox{      if  } 0 < |p| < \frac{4}{5}, \qquad \hbox{  zero otherwise,}\\
\sigma_{h}(p) &=& \sigma(p) \,\,\,\, \hbox{      if  } \frac{4}{5} < |p| < 1, \qquad \hbox{  zero otherwise,}
\end{eqnarray}
and the other fields are also split similarly
\footnote{For numerical efficiency, we choose a division that is split between $\sigma_{l}$ and $\sigma_{h}$ as $p = \frac{4}{5}$.}.
We choose the renormalization conditions as follows:
\begin{eqnarray}
\Gamma^{(2)}_{\psi} (0, 0) &=& -m_{R},\\
\Gamma^{(2)}_{\sigma} (0, 0) &=& -N_{R},\\
\Gamma^{(3)} (0,0,0) &=& -g_{R},
\end{eqnarray}
where $\Gamma^{(i)}$ are renormalized $i$-point functions, $m_{R}, N_{R}, g_{R}$ are renormalized parameters, and the arguments of $\Gamma^{(i)}$ are external momenta. 
In order to obtain effective parameters, we calculate the one-loop effect and integrate out only high-frequency modes:
\begin{eqnarray}
m_{R\alpha\beta} &=& \left(\frac{5}{4}\right)^{-2} \eta_{\psi}^{2} \left[ m -  g^{2} \int_{\frac{4}{5}< |k| <1 } \frac{d^{2}k}{(2 \pi)^{2}} S_{\alpha\beta}(k) D(k)  \right]  , \\ \label{eq.m_pr}
\frac{N_{R}}{2} &=& \left(\frac{5}{4}\right)^{-2} \eta_{\sigma}^{2}\left[ \frac{N}{2} + \frac{g^{2}}{2} \int_{\frac{4}{5}< |k| <1 } \frac{d^{2}k}{(2 \pi)^{2}} \mathrm{tr} \left[ S(k) S(k) \right] \right] , \\ \label{eq.N_pr}
g_{R} &=& \left(\frac{5}{4}\right)^{-4} \eta_{\psi}^{2} \eta_{\sigma} \left[ g + g^{3}  \int_{\frac{4}{5}< |k| <1 } \frac{d^{2}k}{(2 \pi)^{2}} (S(k) S(k))_{\alpha\beta} D(k) \right] \cdot \delta_{\alpha\beta},\label{eq.g_pr} \nonumber \\
 & &
\end{eqnarray}
where $S(k)$ and $D(k)$ are propagators of each field presented below, 
``$\mathrm{tr}$" is a trace operation of the fermionic indices, 
and $\eta_{\psi}$ and $\eta_{\sigma}$ are rescaling parameters for the fermion and auxiliary field respectively.
We can define these parameters with dimensional analysis in the following values:
\begin{eqnarray}
\eta_{\psi} &=& \left( \frac{5}{4} \right)^{3/2} ,\\
\eta_{\sigma} &=&  \frac{5}{4} .
\end{eqnarray}
We can obtain propagators from the GN action:
\begin{eqnarray}
S(k) &=& \left[ D_{f}(\tilde{k}+k) + m \right]^{-1},  \label{eq.2p_psi} \\
D(k) &=& \frac{1}{N},\label{eq.2p_sigma}
\end{eqnarray}
where $D_{f}(k)$ is one of the lattice fermion kinetic terms in Eq.(\ref{eq.D(p)}) and $\tilde{k}$ is the zero-mode momentum in Eq.(\ref{eq.pole}).
Substituting Eqs.(\ref{eq.2p_psi}) and (\ref{eq.2p_sigma}) for Eqs. (\ref{eq.m_pr})--(\ref{eq.g_pr}), we can obtain the effective mass and coupling constant after integrating out over fields $\sigma(\frac{4}{5} < |k| < 1), \psi(\frac{4}{5} < |k| < 1)$, and $\bar{\psi}(\frac{4}{5} < |k| < 1)$.

\end{document}